\documentclass{PoS}

\title{Confinement: $G_2$ group case}

\ShortTitle{Confinement: $G_2$ group case}

\author{\speaker{Guido Cossu}\\
        Scuola Normale Superiore \& INFN, Pisa, Italy\\
        E-mail: \email{g.cossu@sns.it}}
\author{Massimo D'Elia\\
        Dipartimento di Fisica \& INFN, Genova, Italy\\
        E-mail: \email{delia@ge.infn.it}}
\author{Adriano Di Giacomo\\
        Dipartimento di Fisica \& INFN, Pisa, Italy\\
        E-mail: \email{digiacomo@df.unipi.it}}
\author{Biagio Lucini\\
        Department of Physics, Swansea University, Swansea, UK\\
        E-mail: \email{b.lucini@swansea.ac.uk}}
\author{Claudio Pica\\
        Physics Department, Brookhaven National Laboratory, Upton, NY 11973, USA\\
        E-mail: \email{pica@bnl.gov}}

\abstract{The gauge group being centreless, $G_2$ gauge theory is a good laboratory for studying the role of the centre of the group for colour confinement in Yang-Mills gauge theories. In this paper, we investigate $G_2$ pure gauge theory at finite temperature on the lattice. By studying the finite size scaling of the plaquette, the Polyakov loop and their susceptibilities, we show that a deconfinement phase transition takes place. The analysis of the pseudocritical exponents give strong evidence of the deconfinement transition being first order. Implications of our findings for scenarios of colour confinement are discussed.}

\FullConference{The XXV International Symposium on Lattice Field Theory\\
		 July 30-4 August 2007\\
		 Regensburg, Germany}

\begin{document}

\section{Introduction and Motivation}
Quark confinement is one of the oldest issues in non perturbative QCD. The fundamental problem is the characterization of the degrees of freedom responsible of confinement. Many choices are investigated in literature, among them center vortices and abelian monopoles (both proposed by 't Hooft) are the most popular. The connection between centre symmetry breaking and deconfinement transition in $SU(N)$ theories, and the Svetitsky-Yaffe conjecture \cite{Svetitsky:1982gs} relating the universality class of the deconfinement transition of a gauge theory to that of one with its centre as a spin sistem in one lower dimension, seem to link the presence of a non trivial centre to confinement. It's therefore an interesting question whether centerless groups present or not a confinement-deconfinement phase transition. One example in this direction is the $SO(3) = SU(2)/Z_2$ that has been extensively studied on the lattice (see for example \cite{Datta:1999np, Barresi:2004qa, deForcrand:2002vs}). Another choice is the exceptional group $G_2$ which moreover is without 't Hooft center vortices and is the simplest one with this property, so we focused on it for our investigations.

Another point of view concerning the confinement mechanism is the so called \emph{dual superconductor picture}, the subject of the project. The spontaneous breaking of the electromagnetic symmetry to a discrete group in ordinary superconductors, due to condensation of Cooper pairs, leads to the formation of the so called Abrikosov flux tubes connecting magnetic charges. The dual mechanism should connect quark pairs by strings of (chromo)electric flux tubes due to condensation of magnetic objects. According to this Dual Superconductor Picture (DSP), the QCD vacuum,  is a condensate of magnetically charged fields (monopoles) confining (chromo)electrically charged particles. The $G_2$ group admits stable monopole solutions, so in principle there could be a phase where these condense giving rise to confinement even without the presence of a centre.

 We started our investigation by analyzing in detail the thermodynamical properties of the exceptional group $G_2$ and we report here our findings. The first and essential issue is to check whether the transition found in \cite{Pepe:2005sz} is a real one or a crossover, for example. Were this the case, the physics of deconfinement in $G_2$ would be noticeably different from that of $SU(N)$ gauge theories, and this would cast serious doubts about what can we learn from $G_2$ for confinement in more physical gauge theories. 

We perform a Finite Size Scaling (FSS) analysis of the plaquette and of the Polyakov loop susceptibilities. We test the educated guess of a first order phase transition and we can show that a real transition takes place so the confined and deconfined phase are really different. In section \ref{G_2} we discuss some basic properties of the $G_2$ group. Then, in section \ref{simulations} we state the  results of our lattice simulations. We draw the conclusions in the last section.

\subsection{$G_2$ group}
\label{G_2}

We are now going to state some basic facts about the Lie Group $G_2$. It can be naturally constructed as a subgroup of the real group $SO(7)$ which has 21 generators and rank 3. To the usual properties of $SO(7)$ matrices
\begin{equation}
  \det \Omega = 1 \qquad \Omega^{-1} = \Omega^{T}
\end{equation}
we have in addition another constraint
\begin{equation}
  T_{abc} = T_{def} \Omega_{da} \Omega_{eb} \Omega_{fc}
  \label{constraint}
\end{equation}
where $T_{abc}$ is a totally antisymmetric tensor whose nonzero elements are (using the octonion basis given by \cite{Cacciatori:2005yb})
\begin{equation}
  T_{123} = T_{176} = T_{145} = T_{257} = T_{246} = T_{347} = T_{365} = 1.
\end{equation}
Equations \ref{constraint} are 7 relations reducing the numbers of generators to 14. The fundamental representation of $G_2$ is 7 dimensional and by using the algebra representation of \cite{Cacciatori:2005yb} we can clearly identify an $SU(3)$ subgroup and several $SU(2)$ subgroups, with 6 of them we can cover the whole group. The first three $SU(2)$ subgroups are the $4 \times 4$ real representations of the group while the remaining three are extremely difficult to simulate with standard techniques. See next section for details on simulations.

The Lie group $G_2$ has rank 2 as $SU(3)$, this implies that its Cartan subgroup, that is the maximal residual abelian subgroup after an abelian gauge fixing, is $U(1) \times U(1)$. Stable monopole solutions are classified according to the homotopy group\footnote{The first equality follows from $\pi_1(G_2) = 0$. See for example \cite{Weinberg:1996kr}.}:
\begin{equation}
  \pi_2(G_2/U(1)^2) = \pi_1(U(1) \times U(1)) = \mathbb Z \times \mathbb Z 
\end{equation}
i.e. we have two distinct species of monopoles like in $SU(3)$. 

Another interesting homotopy group shows that center vortices are absent in the theory:
\begin{equation}
  \pi_1(G_2/\mathcal C(G_2)) = \pi_1(G_2) = 0
\end{equation}
while for $SU(3)$ for example
\begin{equation}
  \pi_1(SU(3)/\mathbb Z_3) = \mathbb Z_3.
\end{equation}
So $G_2$ is a good playground to study the dual superconductor picture in a theory without center vortices, thus isolating monopole contribution in confinement.

\section{Simulation and Results}
\label{simulations}
Much work has been devoted to simulations of $G_2$ Yang-Mills theory by the Bern group \cite{Holland:2003jy, Pepe:2005sz}. Here we are going to investigate in some detail the thermodynamical properties of the system to demonstrate that a deconfinement phase transition is indeed taking place.

To simulate the gauge theory
\begin{equation}
\mathcal L = \frac{1}{7g^2} {\rm Tr} \, F_{\mu\nu}F_{\mu\nu}
\end{equation}
discretized with the Wilson action, we used a simple Cabibbo-Marinari update (heat-bath + overrelaxation in a tunable ratio) for the first three $SU(2)$ subgroups ($4 \times 4$ representations) spanning the $SU(3) \subset G_2$. To guarantee the covering of the whole gauge group we make a completely random gauge transformation every $n$ updates (tipically 2). The observables measured are the standard plaquette and the Polyakov Loop. To study the thermodynamical properties we simulated several lattices of spatial dimension $N_s = 12, 16, 18, 20$ and $N_t = 6$ ($N_t = 4$ only for the smallest lattice). At the transition we needed histories of the order of $10^5$ updates. The code uses only real algebra and run on an Opteron farm here in the computer facilities of the Physics Department of the University of Pisa. 

\subsection{Thermodynamics}

\begin{figure}
\includegraphics[width=.49\textwidth]{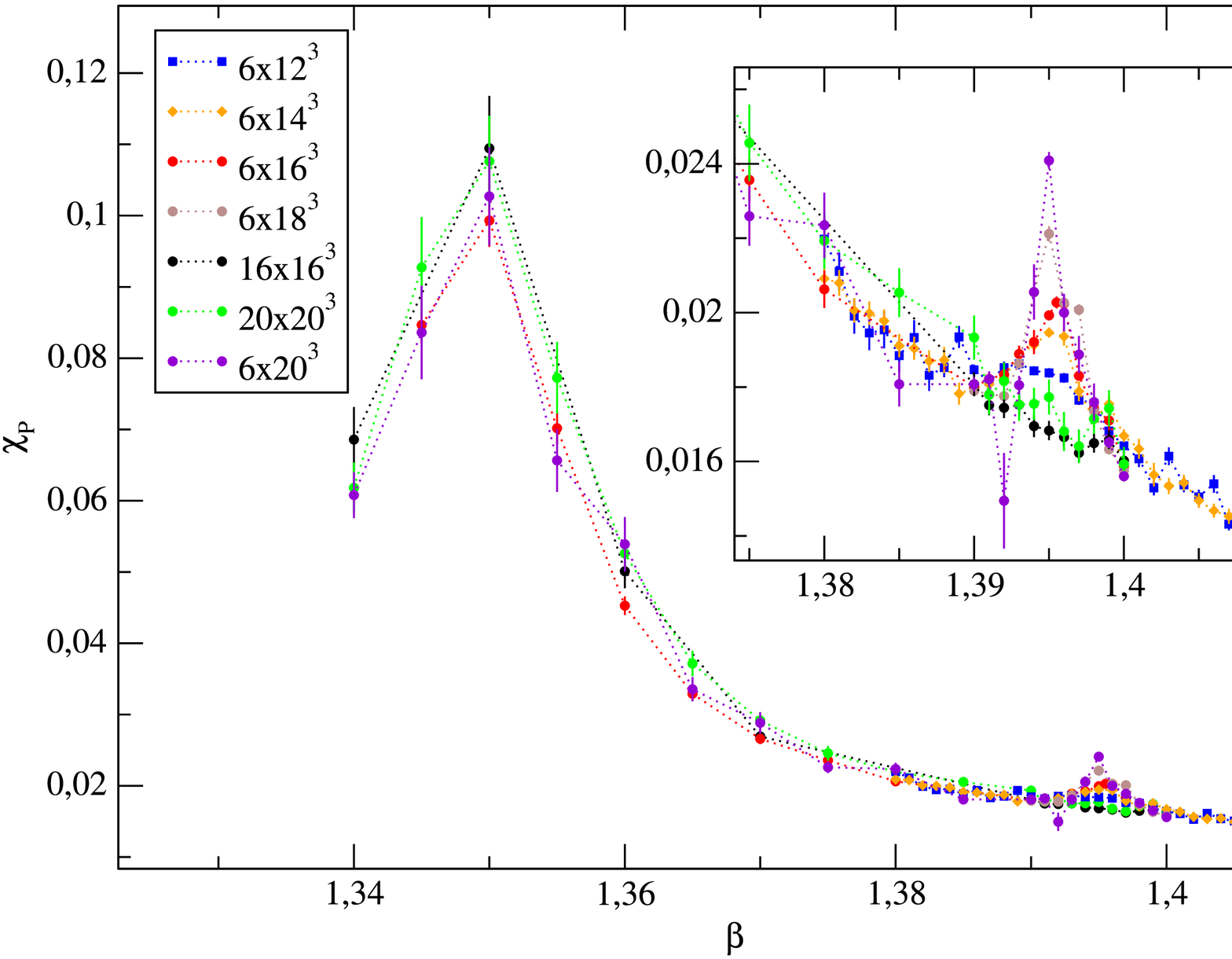}
\includegraphics[width=.49\textwidth]{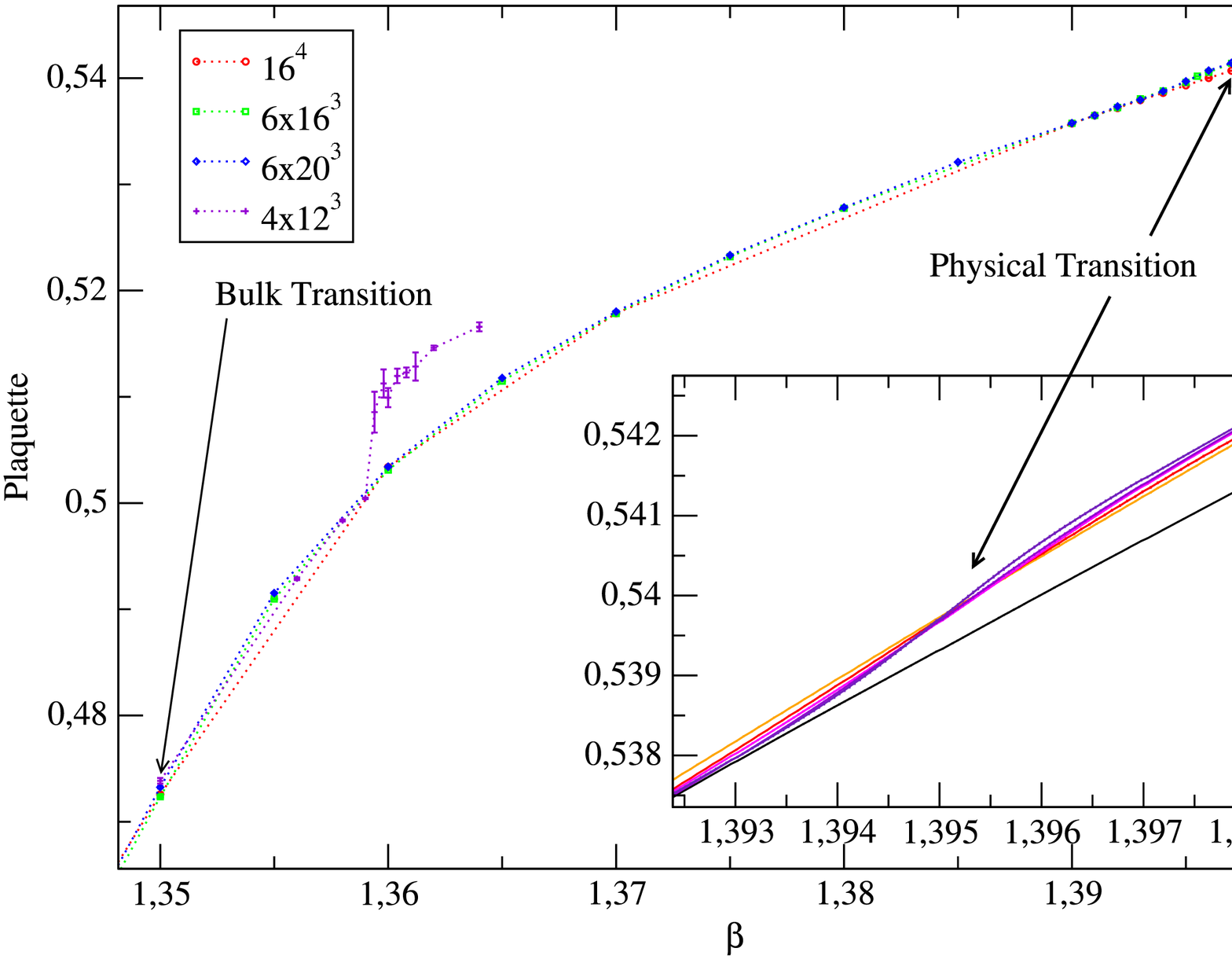}
\caption{Left: plaquette susceptibility for various spatial volumes and $N_t$. The huge peak at $\beta=1.35$ is the bulk transition described in the text. In the inset is shown the real phase transition point. Right: comparison of plaquette for non zero and zero temperature simulations. The integral of the difference of the curves gives the free energy density which is different from zero only at the physical transition.}
\label{Suscplaq}
\end{figure}

\begin{figure}[h]
\includegraphics[width=.51\textwidth]{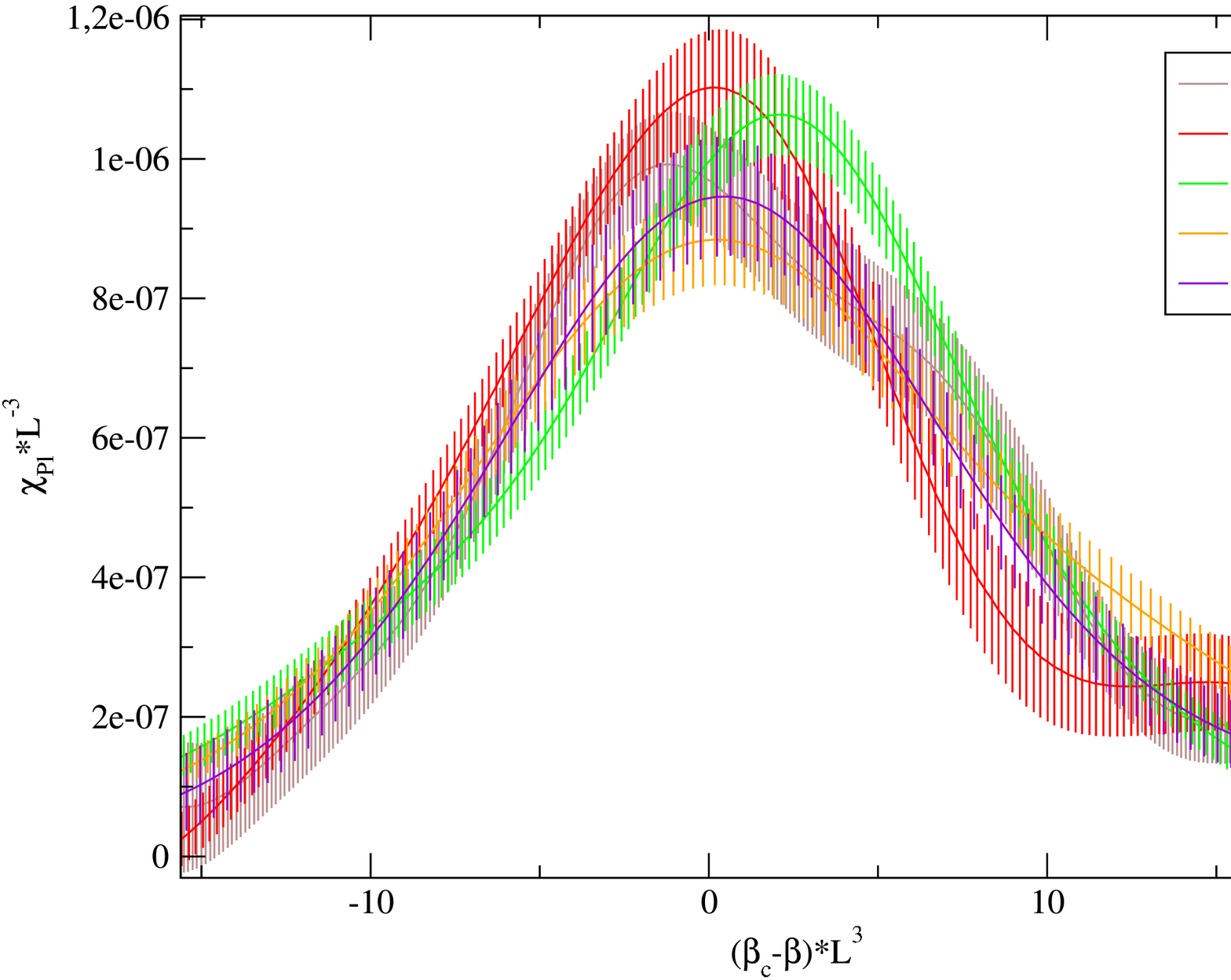}
\includegraphics[width=.48\textwidth]{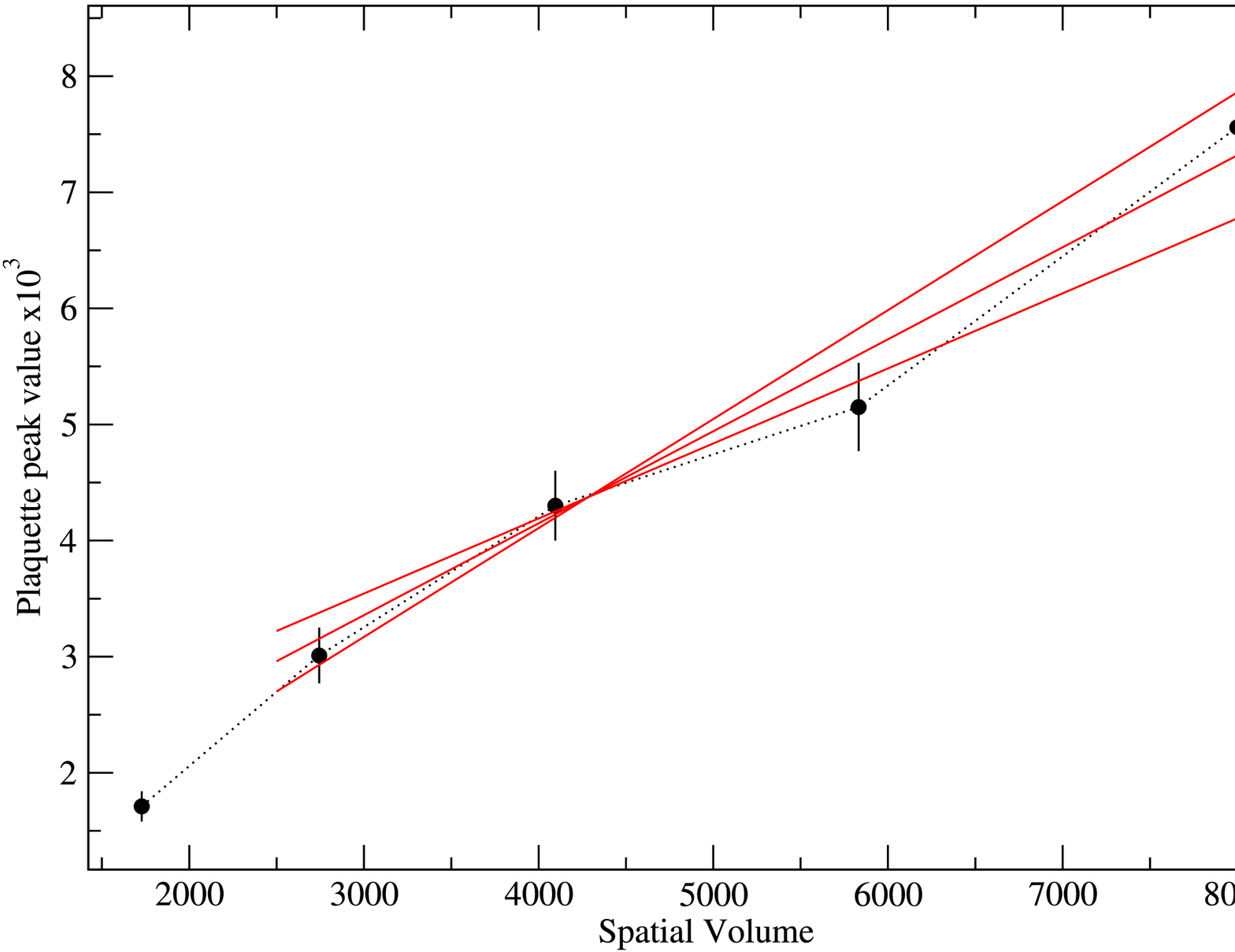}
\caption{Left: scaling of the plaquette susceptibility with a first order transition hypotesis after the subtraction of the unphysical background. Right: Linear fit of the peak heights, $y = a\cdot x + b$, $a = 0.00079(14)\cdot 10^{-3}$, $b = 0.98(62)\cdot10^{-3}$, $\chi^2_{red} = 1.35$.}
\label{scaling}
\end{figure}

The measured observables are the plaquette and the Polyakov loop, and their susceptibilities. The lattice specific heat has a strong peak at  $\beta \simeq 1.35$. This does not signal a real phase transition. It is present for every volume and different $N_t$ (a physical transition should move to the weak coupling region with $N_t$) and also at zero temperature simulations and can be ascribed to lattice artifacts (see Fig. \ref{Suscplaq}).
This big peak completely overshadows the real physical transition that can be seen as a little peak in the weak coupling region at $\beta \sim 1.395$ for $N_t = 6$ and several spatial volumes. We need a subtraction of this background to study the finite size scaling of the plaquette susceptibility. We estimate the background by mean of zero temperature simulations at different spatial volumes (to keep under control possible systematic errors). The scaling of the susceptibility assuming a first order phase transition is shown in Fig. \ref{scaling}. The specific heat scales nicely as it is confirmed by the linear fit of the behaviour of the plaquette susceptibility peak with the volume 

\begin{figure}
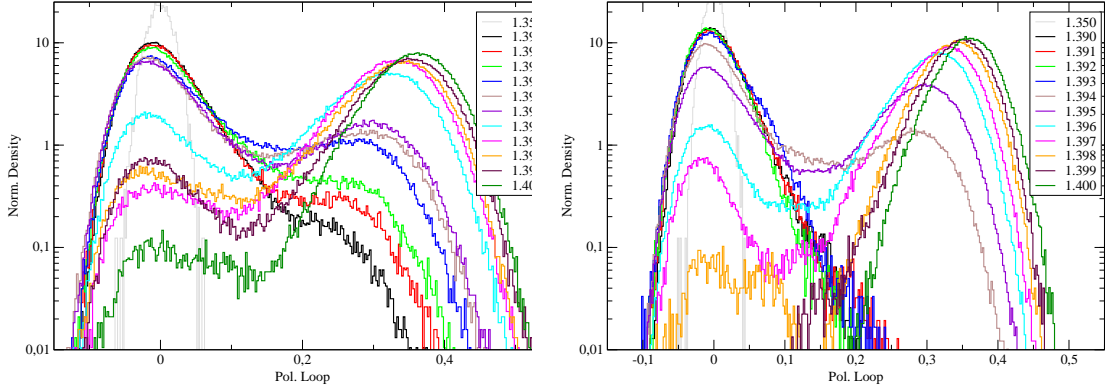

\includegraphics[width=.48\textwidth]{Polyakov_Loop_Density_16}
\includegraphics[width=.48\textwidth]{Polyakov_Loop_Density_20}
\caption{Normalized densities of the Polyakov Loop in a semilog plot for
    $\beta$ varying in the range from 1.35, the critical coupling of the bulk
    transition ``$\beta_{bulk}$'', to 1.401, in the deconfined phase (data
    from the $6\times16^3$ lattice for the upper graph and from $6\times20^3$
    for the other - same scales and limits for both axes are used for better
    comparison). As an aside we notice that far in the confined phase,
    $\beta_c < 1.395$, the Polyakov loop is zero within errors and this feature can not be explained on the ground of any manifest symmetry of the system. }
\label{PolLoopDens}
\end{figure}

The Polyakov loop is insensitive to the bulk transition. We studied its FSS even if it is not a proper order parameter. It has anyway a divergence at the transition, signal of a non trivial overlap with the real (unknown) order parameter, so that we can safely extract the critical exponents. It develops an evident double peak structure (see Fig. \ref{PolLoopDens}). 
From the behaviour of this observable one can guess a first order transition. A  scaling analysis shows clearly  the expected behaviour for large volumes, confirming the results of the specific heat.
\begin{figure}[b]
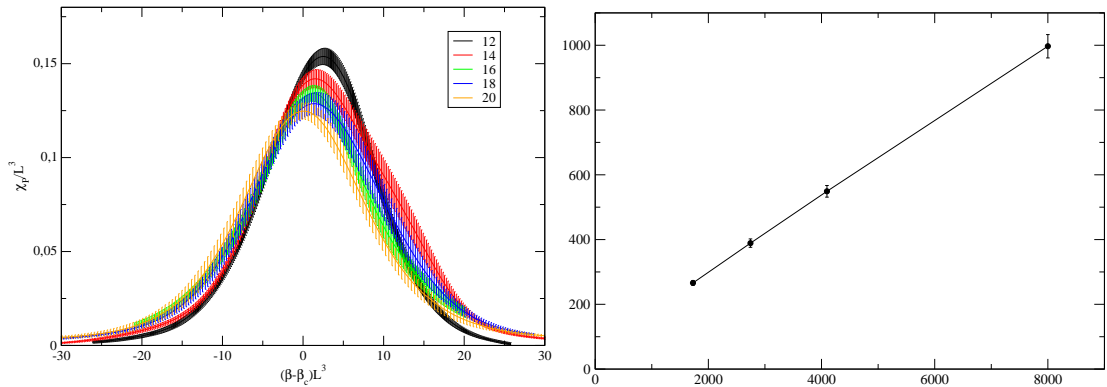

\includegraphics[width=.48\textwidth,clip=]{Polyakov_Loop_Rescaling}
\includegraphics[width=.48\textwidth,clip=]{Polyakov_Loop_Peak_Scaling}
\caption{Left: Scaling of the Polyakov loop assuming first order. For the smallest lattices corrections to the scaling are evident; $\beta_c = 1.395$. Right: Scaling of the peak of the Polyakov Loop susceptibility. The solid line is a linear fit to the data.}
\label{PLscaling}
\end{figure}

\subsection{Conclusions}

We simulated the $G_2$ Yang-Mills theory at finite temperature. In order to investigate the properties of the vacuum by mean of a magnetic charged operator we need to confirm that a real confinement-deconfinement transition is taking place. We showed that the specific heat and the Polyakov Loop observables agree with the hypotesis of a first order transition after subtraction of a background due to lattice artifacts. The centre of the group seems an unessential property for the presence of a deconfinement transition, at least in this case. Future developements concern the measure of the magnetic order parameter proposed by the Pisa group to study the topological structure of the vacuum of this centreless theory.

The work of C.P. has been supported in part by contract DE-AC02-98CH1-886 with the U.S. Department of Energy.

\end{document}